\documentclass{article}
\usepackage{amssymb}
\usepackage{amsfonts}
\usepackage{amsmath}

\input{tcilatex}

\begin{document}

\title{Elastic Plates Motions with Transverse Variation of Microrotation. }
\author{Lev Steinberg \\
%EndAName
Department of Mathematical Sciences\\
University of Puerto Rico\\
Mayaguez, PR 00681-9018, USA}
\maketitle

\begin{abstract}
The purpose of this paper is to present a new mathematical model for the
dynamics of thin Cosserat elastic plates. Our approach, which is based on a
generalization of the classical Reissner-Mindlin plate theory, takes into
account the transverse variation of microrotation and corresponding
microintertia of the the elastic plates. The model assumes polynomial
approximations\ over the plate thickness of asymmetric stress, couple
stress, displacement, and microrotation, which are consistent with the
elastic equilibrium, boundary conditions and the constitutive relationships.
Based on the generalized Hellinger-Prange-Reissner variational principle for
the dynamics and strain-displacement relation we obtain the complete dynamic
theory of Cosserat plate.

\smallskip


\end{abstract}

\section{Introduction}

This paper is straightforward extension of the static theory of Cosserat
plates \cite{Steinberg} for the dynamic case. In order to describe dynamics
of elastic plates with microstructure that possess grains, particles,
fibers, and cellular structures \ A. C. Eringen (1967) was the first to
propose a theory of plates in the framework of Cosserat (micropolar)
Elasticity \cite{Eringen}. His theory is based on a direct technique of
integration of the Cosserat Elasticity and assumes no variation of
micropolar rotations in the thickness direction. Eringen's plate theory in
the current form does not produce the Reissner-Mindlin plate equations for
zero microrotations. In this paper we propose to use the Reissner-Mindlin's
plate theory as a foundation for the modeling of dynamics Cosserat elastic
plates. Our approach, in addition to the transverse shear deformation, takes
into account the second order approximation of couple stresses and the
variation of micropolar rotations in the thickness direction and the
corresponding inertia characteristics

\section{Micropolar (Cosserat) Linear Elasticity}

\subsection{Fundamental Equations}

Before proceeding some notation convention should be explained. We use the
usual summation conventions and all expressions that contain Latin letters
as subindices are understood to take values in the set $\{1,2,3\}$. When
Greek letters appear as subindices then it will be assumed that they can
take the values $1$ or $2$.

The Cosserat elasticity equilibrium equations without body forces represent
the balance of linear and angular momentums of micropolar elasticity and
have the following form \cite{Eringen}: 
\begin{eqnarray}
\mathbf{div\sigma } &\mathbf{=}&\rho \mathbf{\ddot{u}},
\label{equilibrium_equations 0} \\
\mathbf{\varepsilon \cdot \sigma }\text{ +}\mathbf{div\mu } &\mathbf{=}&%
\mathbf{J\ddot{\varphi}},  \label{equilibrium_equations}
\end{eqnarray}%
{where} the quantity $\mathbf{\sigma =}\left\{ \sigma _{ji}\right\} $ is the
stress tensor, $\mathbf{\mu =}\left\{ \mu _{ji}\right\} $ the couple stress
tensor, $\mathbf{u}$ and \textbf{\ }$\mathbf{\varphi }$ \ the displacement
and rotation vectors, $\rho $ and $\mathbf{J=}\left\{ J_{i}\right\} $ the
material density and the rotatory inertia characteristics, and $\mathbf{%
\varepsilon }=\left\{ \varepsilon _{ijk}\right\} $ is the Levi--Civita
tensor, where $\varepsilon _{ijk}$ equals 1 or -1 according as $(i,j,k)$ is
an even or odd permutation of 1,2,3 and zero otherwise, and $\mathbf{%
\varepsilon \cdot \sigma =}\left\{ \mathbf{\varepsilon }_{ijk}\sigma
_{jk}\right\} .$The constitutive equation can be written in Nowacki's form 
\cite{Nowaki}:

\begin{eqnarray}
\mathbf{\sigma } &=&(\mu +\alpha )\mathbf{\gamma }+(\mu -\alpha )\mathbf{%
\gamma }^{T}+\lambda \mathbf{(tr\gamma )1},  \label{Hooke's_law 1} \\
\mathbf{\mu } &=&(\gamma +\epsilon )\mathbf{\chi }+(\gamma -\epsilon )%
\mathbf{\chi }^{T}+\beta \mathbf{(tr\chi )1},  \label{Hooke's_law 1A}
\end{eqnarray}%
which we consider with{\ the strain-displacement and torsion-rotation
relations}

\begin{equation}
\mathbf{\gamma =}\left( \mathbf{\nabla u}\right) ^{T}\mathbf{+\varepsilon
\cdot \varphi }\text{ and }\mathbf{\chi }=\nabla \mathbf{\varphi },
\label{kinematic formulas}
\end{equation}%
where quantities $\mathbf{\gamma }$ and $\mathbf{\chi }$ $,$ are the
micropolar strain and torsion tensors, $\mathbf{1}$ the identity tensor, $%
\mathbf{\mu },\lambda $ are the symmetric and $\beta ,\mathbf{,}\gamma 
\mathbf{,}\epsilon ,\alpha $ \ the asymmetric Cosserat elasticity constants

The constitutive equations can be written in the reversible form:%
\begin{eqnarray}
\mathbf{\gamma } &\mathbf{=}&(\mu ^{\prime }+\alpha ^{\prime })\mathbf{%
\sigma }+(\mu ^{\prime }-\alpha ^{\prime })\mathbf{\sigma }^{T}+\lambda
^{\prime }\mathbf{(tr\sigma )1},  \label{Hooke's_law 2} \\
\mathbf{\chi } &\mathbf{=}&(\gamma ^{\prime }+\epsilon ^{\prime })\mathbf{%
\mu }+(\gamma ^{\prime }-\epsilon ^{\prime })\mathbf{\mu }^{T}+\beta \mathbf{%
(tr\mu )1}.  \label{Hooke's_law 2A}
\end{eqnarray}%
where{\ }$\mu ^{\prime }=\frac{1}{4\mu }$, $\alpha ^{\prime }=\frac{1}{%
4\alpha }$, $\gamma ^{\prime }=\frac{1}{4\gamma }$, $\epsilon ^{\prime }=%
\frac{1}{4\epsilon }$, $\lambda ^{\prime }=\frac{-\lambda }{6\mu (\lambda +%
\frac{2\mu }{3})}$ and $\beta ^{\prime }=\frac{-\beta }{6\mu (\beta +\frac{%
2\gamma }{3})}$.\footnote{%
Neff \cite{Neff 1} uses the notation $\mu _{c}$ for elastic parameter $%
\alpha .$}

We consider a Cosserat elastic body $B_{0}.$ In this case the equilibrium
equations (\ref{equilibrium_equations 0}) - (\ref{equilibrium_equations})
with constitutive formulas (\ref{Hooke's_law 1}) - (\ref{Hooke's_law 1A})
and kinematics formulas (\ref{kinematic formulas}) should be accompanied by
the following mixed boundary%
\begin{eqnarray}
\mathbf{u} &\mathbf{=}&\mathbf{u}_{o}\mathbf{,}\text{ }\mathbf{\varphi
=\varphi }_{o}\text{ on }\mathcal{G}_{1}=\partial B_{0}\backslash \partial
B_{\sigma },  \label{Bound conditions} \\
\mathbf{\sigma }_{\mathbf{n}} &=&\mathbf{\sigma \cdot n}=\mathbf{\sigma }%
_{o},\text{ }\mathbf{\mu }_{\mathbf{n}}=\mathbf{\mu }\cdot \mathbf{n=\mu }%
_{o}\text{ on }\mathcal{G}_{2}=\partial B_{\sigma },
\label{Bound conditions a}
\end{eqnarray}%
and initial conditions%
\begin{eqnarray}
\mathbf{u(x,o)} &\mathbf{=}&\mathbf{U}_{o}\mathbf{,}\text{ }\mathbf{\varphi
(x,o)=\Phi }_{o}\text{ in }B_{0}, \\
\mathbf{\dot{u}(x,o)} &\mathbf{=}&\mathbf{\mathbf{\tilde{U}}_{o},}\text{ }%
\mathbf{\dot{\varphi}(x,o)=\tilde{\Phi}}_{o}\text{ on }\mathcal{B}_{o},
\end{eqnarray}%
where $\mathbf{u}_{o}$\textbf{, }$\mathbf{\varphi }_{o}$\textbf{\ }are
prescribed on $\mathcal{G}_{1}$, $\sigma _{o}$\textbf{\ }and \ $\mathbf{\mu }
$ \ on $\mathcal{G}_{2},$ and $\mathbf{n}$\textbf{\ }denotes the outward
unit normal vector to $\partial B_{0}.$

\subsection{Cosserat Elastic Energy}

The strain stored energy $U_{C}$ of the body $B_{0}$ is defined by the
integral \cite{Nowaki}:

\begin{equation}
{\large U}_{C}=\int_{B_{0}}\emph{W}\left\{ \mathbf{\gamma ,\chi }\right\} dv,
\label{free_energy_strain}
\end{equation}%
where 
\begin{eqnarray}
\emph{W}\left\{ \mathbf{\gamma ,\chi }\right\} &=&\frac{\mu +\alpha }{2}%
\gamma _{ij}\gamma _{ij}+\frac{\mu -\alpha }{2}\gamma _{ij}\gamma _{ji}+%
\frac{\lambda }{2}\gamma _{kk}\gamma _{nn}  \label{free_energy_strain A} \\
&&+\frac{\gamma +\epsilon }{2}\chi _{ij}\chi _{ij}+\frac{\gamma -\epsilon }{2%
}\chi _{ij}\chi _{ji}+\frac{\beta }{2}\chi _{kk}\chi _{nn},  \notag
\end{eqnarray}%
then the constitutive relations (\ref{Hooke's_law 1}) - (\ref{Hooke's_law 1A}%
) can be written in the form:

\begin{equation}
\mathbf{\sigma }=\mathbf{C}_{\sigma }\left[ \emph{W}\right] =\nabla _{%
\mathbf{\gamma }}\emph{W}\text{ and }\mathbf{\mu }=\mathbf{C}_{\mu }\left[ 
\emph{W}\right] =\nabla _{\mathbf{\chi }}\emph{W}.  \label{Const-d 2}
\end{equation}

The function $\emph{W}$ is non-negative if and only if \cite{Nowaki} 
\begin{eqnarray}
\mu &>&0,\ \ 3\lambda +2\mu >0,  \notag \\
\gamma &>&0,\ \ 3\beta +2\gamma >0,  \label{parameter conditions} \\
\alpha &>&0,\ \ \mu +\alpha >0,  \notag \\
\epsilon &>&0,\ \ \gamma +\epsilon >0.  \notag
\end{eqnarray}%
For future convenience, we present the stress energy%
\begin{equation*}
{\large U}_{K}=\int_{B_{0}}\Phi \left\{ \sigma \mathbf{,\mu }\right\} dv,
\end{equation*}%
where%
\begin{eqnarray}
\Phi \left\{ \sigma \mathbf{,\mu }\right\} &=&\frac{\mu ^{\prime }+\alpha
^{\prime }}{2}\sigma _{ij}\sigma _{ij}+\frac{\mu ^{\prime }-\alpha ^{\prime }%
}{2}\sigma _{ij}\sigma _{ji}+\frac{\lambda ^{\prime }}{2}\sigma _{kk}\sigma
_{nn}  \notag \\
&&+\frac{\gamma ^{\prime }+\epsilon ^{\prime }}{2}\mu _{ij}\mu _{ij}+\frac{%
\gamma ^{\prime }-\epsilon ^{\prime }}{2}\mu _{ij}\mu _{ji}+\frac{\beta
^{\prime }}{2}\mu _{kk}\mu _{nn}.  \label{free_energy_stress}
\end{eqnarray}%
\ The reversible constitutive relation (\ref{Hooke's_law 2}) - (\ref%
{Hooke's_law 2A}) can be also written in form:%
\begin{equation}
\mathbf{\gamma =K}_{\gamma }\left[ \mathbf{\sigma }\right] =\frac{\partial
\Phi }{\partial \mathbf{\sigma }},\text{ }\mathbf{\chi =K}_{\chi }\left[ 
\mathbf{\mu }\right] =\frac{\partial \Phi }{\partial \mathbf{\mu }}.
\label{Const1-1B}
\end{equation}

The total internal work done by the stresses $\mathbf{\sigma }$ and $\mathbf{%
\mu }$ over the strains $\mathbf{\ \gamma }$ and $\mathbf{\chi }$ $\ $for
the body $B_{0}$ \cite{Nowaki} is 
\begin{equation}
{\large U}=\int_{B_{0}}\left[ \mathbf{\sigma \cdot \gamma +\mu \cdot \chi }%
\right] dv  \label{free energy}
\end{equation}%
and 
\begin{equation*}
{\large U=U}_{K}{\large =U}_{C}
\end{equation*}%
provided the constitutive relations (\ref{Hooke's_law 1}) - (\ref%
{Hooke's_law 1A}) hold.

\subsection{The Generalized Hellinger-Prange -Reissner (HPR)\ Principle}

The HPR principle \cite{Gurtin} in the case of Cosserat elasticity\ states,
that for any set $\mathcal{A}$ of all admissible states $\mathfrak{s=}\left[ 
\mathbf{u,\varphi ,\gamma ,\chi ,\sigma ,\mu }\right] $ that satisfy the {%
strain-displacement, torsion-rotation relations (\ref{kinematic formulas})}
and the initial condition{,} the zero variation 
\begin{equation*}
\delta \Theta (\mathfrak{s})=0
\end{equation*}

of the functional 
\begin{eqnarray}
\Theta (\mathfrak{s}) &=&U_{K}-\int_{B_{0}}\left[ \mathbf{\sigma \cdot
\gamma -\rho \ddot{u}\cdot u+\mu \cdot \chi -J}\text{ }\mathbf{\ddot{\varphi}%
\cdot \varphi }\right] dv  \label{Var 2} \\
&&+\int_{\mathcal{G}_{1}}\left[ \mathbf{\sigma }_{\mathbf{n}}\cdot (\mathbf{%
u-u}_{o})+\mathbf{\mu }_{\mathbf{n}}\left( \mathbf{\varphi -\varphi }%
_{o}\right) \right] da+\int_{\mathcal{G}_{2}}\left[ \mathbf{\sigma }%
_{o}\cdot \mathbf{u+m}_{o}\cdot \mathbf{\varphi }\right] da  \notag
\end{eqnarray}%
\bigskip at $\mathfrak{s\in }\mathcal{A}$ is equivalent of $\mathfrak{s}$ to
be a solution of the system of equilibrium equations (\ref%
{equilibrium_equations 0}) - (\ref{equilibrium_equations}), constitutive
relations (\ref{Hooke's_law 2}) - (\ref{Hooke's_law 2A}), which satisfies
the mixed boundary conditions (\ref{Bound conditions}) - (\ref{Bound
conditions a}). The proof is similar to the proof for HPR\ principle for
classic linear elasticity \cite{Gurtin}.

\section{The Cosserat Plate Assumptions}

In this section we formulate our stress, couple stress and kinematic
assumptions of the Cosserat plate. The set of points $P=\left\{ \Gamma
\times \left[ -h/2,h/2\right] \right\} \cup T\cup B$ forms the entire
surface of the plate and $\left\{ \Gamma _{u}\times \left[ -h/2,h/2\right]
\right\} $ is the lateral part of the boundary where displacements and
microrotations are prescribed. The notation $\Gamma _{\sigma }=\Gamma
\backslash \Gamma _{u}$ of the remainder we use to describe the lateral part
of the boundary edge $\left\{ \Gamma _{\sigma }\times \left[ -h/2,h/2\right]
\right\} $ where stress and couple stress are prescribed. We also use
notation $P_{0}$ \ for the middle plane internal domain of the plate.

In our case we consider the vertical load and pure twisting momentum
boundary conditions at the top and bottom of the plate, which can be written
in the form:

\begin{eqnarray}
\sigma _{33}|_{x_{3}=h/2} &=&\sigma ^{t}(x_{1},x_{2},t),\text{ }%
_{33}|_{x_{3}=-h/2}=\sigma ^{b}(x_{1},x_{2},t),  \label{Bound conditions 0}
\\
\text{ \ \ \ \ \ }\sigma _{3\beta }|_{x_{3}=h/2} &=&0,\text{ }\sigma
_{3\beta }|_{x_{3}=-h/2}=0  \label{Bound conditions 1} \\
\text{ \ \ \ }\mu _{33}|_{x_{3}=h/2} &=&\mu ^{t}(x_{1},x_{2},t),\text{ }\mu
_{33}|_{x_{3}=-h/2}=\mu ^{b}(x_{1},x_{2},t),  \label{Bound conditions 1a} \\
\text{\ }\mu _{3\beta }|_{x_{3}=h/2} &=&0=0,\text{ \ }\mu _{3\beta
}|_{x_{3}=-h/2}=0,  \label{Bound conditions 2}
\end{eqnarray}%
where $(x_{1},x_{2})\in P_{0}.$

\subsection{Stress, Couple Stress and Kinematics Assumptions}

Our approach, which is in the spirit of the Reissner's theory of plates \cite%
{Reissner}, assumes that the variation of stress $\sigma _{kl\text{ }}$and
couple stress $\mu _{kl}$ components across the thickness can be represented
by means of polynomials of $x_{3}$.We adapt the expressions for the stress
and couple-stress components in the following form \cite{Steinberg}:

\begin{equation}
\sigma _{\alpha \beta }=n_{\alpha \beta }(x_{1},x_{2},t)+\frac{h}{2}\zeta
_{3}m_{\alpha \beta }(x_{1},x_{2},t),  \label{stress assumption 1}
\end{equation}%
\begin{equation}
\sigma _{3\beta }=q_{\beta }(x_{1},x_{2},t)\left( 1-\zeta _{3}^{2}\right) ,
\label{stress assumption 1a}
\end{equation}%
\begin{equation}
\sigma _{\beta 3}=q_{\beta }^{\ast }(x_{1},x_{2},t)\left( 1-\zeta
_{3}^{2}\right) .  \label{stress assumption 1b}
\end{equation}%
\begin{equation}
\sigma _{33}=-\frac{3}{4}\left( \frac{1}{3}\zeta _{3}^{3}-\zeta _{3}\right)
p+\sigma _{0},  \label{stress assumption 1ca}
\end{equation}%
\begin{equation}
\mu _{\alpha \beta }=\left( 1-\zeta _{3}^{2}\right) r_{\alpha \beta
}(x_{1},x_{2},t).  \label{stress assumption 1d}
\end{equation}%
\begin{equation}
\mu _{\beta 3}=\zeta _{3}s_{\beta }^{\ast }(x_{1},x_{2},t)+m_{\beta }^{\ast
}(x_{1},x_{2},t).  \label{stress assumption 1e}
\end{equation}%
\begin{equation}
\mu _{3\beta }=0.
\end{equation}%
\begin{equation}
\mu _{33}=\zeta _{3}v+t,\   \label{stress assump 3(a)}
\end{equation}%
where $p=\sigma ^{t}(x_{1},x_{2},t)-\sigma ^{b}(x_{1},x_{2},t)$, $\sigma
_{0}=\frac{1}{2}\left( \sigma ^{t}(x_{1},x_{2},t)+\sigma
^{b}(x_{1},x_{2},t)\right) ,$ $v(x_{1},x_{2})=\frac{1}{2}\left( \mu
^{t}(x_{1},x_{2})-\mu ^{b}(x_{1},x_{2})\right) $ and $t(x_{1},x_{2})=\frac{1%
}{2}\left( \mu ^{t}(x_{1},x_{2})+\mu ^{b}(x_{1},x_{2})\right) $. satisfy the
boundary condition requirements. We note that expression (\ref{stress
assumption 1ca}) is identical to the expression of $\sigma _{33}$ given in 
\cite{Reissner} in the case of $\sigma ^{b}=0$.

We also assume displacements $u_{a}$are also distributed linearly over the
thickness of the plate \cite{Eringen}\ and \ that $u_{3}$ \ does not vary
over the thickness of the plate, i.e.

\begin{eqnarray}
u_{\alpha } &=&U_{\alpha }(x_{1},x_{2},t)-\frac{h}{2}\zeta _{3}V_{\alpha
}(x_{1},x_{2},t),  \label{kin1} \\
u_{3} &=&w(x_{1},x_{2},t),  \label{kin1a}
\end{eqnarray}%
where the terms $V_{\alpha }(x_{1},x_{2},t)$ represent the rotations in
middle plane. The variation of microrotation with respect to $x_{3}$ be
represented by means of the second and third order polynomials \cite%
{Steinberg}$:$

\begin{eqnarray}
\varphi _{\alpha } &=&\Theta _{\alpha }^{0}(x_{1},x_{2},t)\left( 1-\zeta
_{3}^{2}\right) ,  \label{kin2a} \\
\varphi _{3} &=&\Theta _{3}^{0}(x_{1},x_{2},t)+\zeta _{3}\left( 1-\frac{1}{3}%
\zeta _{3}^{2}\right) \Theta _{3}(x_{1},x_{2},t).  \label{kin 2}
\end{eqnarray}%
where $\zeta _{3}=\frac{2}{h}x_{3},\ $ $\alpha ,\beta \in \{1,2\}$.

We also assume that initial condition can be presented in the similar form,
so they can be reduced to the form%
\begin{eqnarray}
U_{\alpha }(x_{1},x_{2},0) &=&U_{\alpha }^{0}(x_{1},x_{2}),V_{\alpha
}(x_{1},x_{2},0)=V_{\alpha }^{0}(x_{1},x_{2}),  \label{IN Con 1} \\
w(x_{1},x_{2},0) &=&w^{0}(x_{1},x_{2})
\end{eqnarray}

\begin{eqnarray}
\dot{U}_{\alpha }(x_{1},x_{2},0) &=&\tilde{U}_{\alpha }^{0}(x_{1},x_{2}),%
\dot{V}_{\alpha }(x_{1},x_{2},0)=\tilde{V}_{\alpha }^{0}(x_{1},x_{2}), \\
\dot{w}(x_{1},x_{2},0) &=&\tilde{w}^{0}(x_{1},x_{2})  \notag
\end{eqnarray}%
and 
\begin{eqnarray}
\Theta _{\alpha }^{0}(x_{1},x_{2},0) &=&\Theta _{o\alpha }^{0}(x_{1},x_{2}),
\label{IN Con 2} \\
\Theta _{3}^{0}(x_{1},x_{2},0) &=&\Theta _{o3}^{0}(x_{1},x_{2})\text{ and }%
\Theta _{3}(x_{1},x_{2},t)=\Theta _{o3}(x_{1},x_{2}).  \notag
\end{eqnarray}%
\begin{eqnarray}
\dot{\Theta}_{\alpha }^{0}(x_{1},x_{2},0) &=&\tilde{\Theta}_{o\alpha
}^{0}(x_{1},x_{2}), \\
\dot{\Theta}_{3}^{0}(x_{1},x_{2},0) &=&\tilde{\Theta}_{o3}^{0}(x_{1},x_{2})%
\text{ and }\dot{\Theta}_{3}(x_{1},x_{2},t)=\tilde{\Theta}_{o3}(x_{1},x_{2}).
\notag
\end{eqnarray}

\section{Specification of HPR\ Variational Principle for the Dynamics of
Cosserat Plates}

The HPR\ variational principle for a Cosserat plate is most appropriately
expressed in terms of corresponding integrands calculated across the whole
thickness. We also introduce the weighted characteristics of displacements,
microrotations, strains and stresses of the plate, which will be used to
produce the explicit forms of these integrands.

\subsection{The Cosserat plate stress energy density}

We define the plate stress energy density by the formula;

\begin{equation}
{\large \Phi (}\mathcal{S}{\large )}=\frac{h}{2}\int_{-1}^{1}\Phi \left\{
\sigma \mathbf{,\mu }\right\} d\zeta _{3}.  \label{C and Co expression}
\end{equation}

Taking into account the stress and couple stress assumptions (\ref{stress
assumption 1}) - (\ref{stress assump 3(a)}) and by the integrating $\Phi
\left\{ \sigma \mathbf{,\mu }\right\} $ with respect $\zeta _{3\text{ }}$in $%
[-1,1]$ \ we obtain the explicit plate stress energy density expression in
the form \cite{Steinberg}:

\begin{eqnarray}
{\large \Phi (}\mathcal{S}{\large )} &=&\frac{\lambda +\mu }{2h\mu (3\lambda
+2\mu )}\left[ N_{\alpha \alpha }^{2}+\frac{12}{h^{2}}M_{\alpha \alpha }^{2}%
\right]  \notag \\
&&-\frac{\lambda }{2h\mu (3\lambda +2\mu )}\left[ N_{11}N_{22}+\frac{12}{%
h^{2}}M_{11}M_{22}\right]  \notag \\
&&+\frac{\alpha +\mu }{8h\alpha \mu }\left[ (1-\delta _{\alpha \beta
})\left( N_{\alpha \beta }^{2}+\frac{12}{h^{2}}M_{\alpha \beta }^{2}\right) +%
\frac{6}{5}\left( Q_{\alpha }Q_{\alpha }+Q_{\beta }^{\ast }Q_{\beta }^{\ast
}\right) \right]  \notag \\
&&+\frac{3(\alpha -\mu )}{10h\alpha \mu }\left[ Q_{\alpha }Q_{\alpha }^{\ast
}+\frac{5}{6}N_{12}N_{21}+\frac{10}{h^{2}}M_{12}M_{21}\right] +\frac{%
3\lambda }{5h\mu (3\lambda +2\mu )}Q_{\alpha ,\alpha }^{\ast }M_{\beta \beta
}  \notag \\
&&+\frac{3}{5h\gamma (3\beta +2\gamma )}\left[ (\beta +\gamma )R_{\alpha
\alpha }^{2}-\beta R_{11}R_{22}\right] +\frac{3}{10h}\left( \frac{1}{\gamma }%
-\frac{1}{\epsilon }\right) R_{12}R_{21}  \notag \\
&&+\frac{17h(\lambda +\mu )}{280\mu (3\lambda +2\mu )}\left( Q_{\alpha
,\alpha }^{\ast }\right) ^{2}+\frac{\lambda }{2\mu (3\lambda +2\mu )}\left(
N_{\alpha \alpha }\right) \sigma _{0}  \notag \\
&&+\frac{h(\lambda +\mu )}{2\mu (3\lambda +2\mu )}\sigma _{0}^{2}-\frac{%
\gamma +\epsilon }{h\gamma \epsilon }\left[ \frac{1}{8}M_{\alpha }^{\ast
}M_{\alpha }^{\ast }+\frac{3}{2h^{2}}S_{\alpha }^{\ast }S_{\alpha }^{\ast }+%
\frac{3}{20}(1-\delta _{\beta \gamma })R_{\beta \gamma }^{2}\right]  \notag
\\
&&-\frac{\beta }{2\gamma (3\beta +2\gamma )}R_{\alpha \alpha }t+\frac{%
h(\beta +\gamma )}{2\gamma (3\beta +2\gamma )}t^{2}+\frac{h(\beta +\gamma )}{%
6\gamma (3\beta +2\gamma )}v^{2},  \label{Energy density}
\end{eqnarray}%
{where the Cosserat stress set }%
\begin{equation}
\mathcal{S}=\left[ M_{\alpha \beta },Q_{\alpha },Q_{3\alpha }^{\ast
},R_{\alpha \beta },S_{\beta }^{\ast },N_{\alpha \beta },M_{\alpha }^{\ast }%
\right] ,  \label{S1}
\end{equation}%
where

\begin{eqnarray}
M_{\alpha \beta } &=&\left( \frac{h}{2}\right) ^{2}\int_{-1}^{1}\zeta
_{3}\sigma _{\alpha \beta }d\zeta _{3}=\frac{h^{3}}{12}m_{\alpha \beta },
\label{Stress resultant 1} \\
Q_{\alpha } &=&\frac{h}{2}\int_{-1}^{1}\sigma _{3\alpha }d\zeta _{3}=\frac{2h%
}{3}q_{\alpha },\text{ }Q_{\alpha }^{\ast }=\frac{h}{2}\int_{-1}^{1}\sigma
_{\alpha 3}d\zeta _{3}=\frac{2h}{3}q_{\alpha }^{\ast }  \notag \\
R_{\alpha \beta } &=&\frac{h}{2}\int_{-1}^{1}\mu _{\alpha \beta }d\zeta _{3}=%
\frac{2h}{3}r_{\alpha \beta },  \notag \\
S_{\alpha }^{\ast } &=&\left( \frac{h}{2}\right) ^{2}\int_{-1}^{1}\zeta
_{3}\mu _{\alpha 3}d\zeta _{3}=\frac{h^{2}}{6}s_{\alpha }^{\ast },  \notag \\
N_{\alpha \beta } &=&\frac{h}{2}\int_{-1}^{1}\sigma _{\alpha \beta }d\zeta
_{3}=hn_{\alpha \beta },\text{ }M_{\alpha }^{\ast }=\frac{h}{2}%
\int_{-1}^{1}\mu _{\alpha 3}d\zeta _{3}=hm_{\alpha }^{\ast },  \notag
\end{eqnarray}

Here $M_{11}$ and $M_{22}$ are the bending moments, $M_{12}$ and $M_{21}$
the twisting moments, $Q_{\alpha }$ the shear forces, $Q_{\alpha }^{\ast }$
the transverse shear forces, $R_{11}$ and $R_{22}$ the micropolar bending
moments, $R_{12}$ and $R_{21}$ the micropolar twisting moments, $S_{\alpha
}^{\ast }$ the micropolar couple moments, all defined per unit length, $%
N_{11}$ and $N_{22}$ are the bending forces, $N_{12}$ and $N_{21}$ the
twisting forces, $M_{\alpha }^{\ast }$ the micropolar shear couple-stress
resultants.

Then the stress energy of the plate $P$%
\begin{equation}
U_{K}^{\mathcal{S}}=\int_{P_{0}}{\large \Phi (}\mathcal{S}{\large )}da,
\label{energy 1}
\end{equation}%
where $P_{0\text{ }}$ is the internal domain of the middle plane of the
plate $P.$

\subsection{The density of the work done over the Cosserat plate boundary}

In the following consideration we also assume that the proposed stress,
couple stress, and kinematic assumptions are valid for the lateral\ boundary
of the plate $P$ as well.

We evaluate the density of the work over the boundary $\Gamma _{u}\times %
\left[ -h/2,h/2\right] $

\begin{equation}
\mathcal{W}_{1}=\frac{h}{2}\int_{-1}^{1}\left[ \mathbf{\sigma }_{\mathbf{n}%
}\cdot \mathbf{u}+\mathbf{\mu }_{\mathbf{n}}\mathbf{\varphi }\right] d\zeta
_{3}.  \label{work 1}
\end{equation}%
Taking into account the stress and couple stress assumptions (\ref{stress
assumption 1}) - (\ref{stress assump 3(a)}) and kinematic assumptions (\ref%
{kin1}) - (\ref{kin 2}) we are able to represent $\mathcal{W}_{1}$ by the
following expression \cite{Steinberg}: 
\begin{equation}
\mathcal{W}_{1}=\mathcal{S}_{n}\mathcal{\cdot U=}\check{M}_{\alpha }\Psi
_{\alpha }+\check{Q}^{\ast }W+\check{R}_{\alpha }\Omega _{\alpha }^{0}+%
\check{S}^{\ast }\Omega _{3}+\check{N}_{\alpha }U_{\alpha }+\check{M}^{\ast
}\Omega _{3}^{0},
\end{equation}%
where the sets $\mathcal{S}_{n}\mathcal{\ }$and $\mathcal{U}$ are defined as%
\begin{eqnarray*}
\mathcal{S}_{n} &=&\left[ \check{M}_{\alpha },\check{Q}^{\ast },\check{R}%
_{\alpha },\check{S}^{\ast },\check{N}_{\alpha },\check{M}^{\ast }\right] ,
\\
\mathcal{U} &=&\left[ \Psi _{\alpha },W,\Omega _{\alpha }^{0},\Omega
_{3},U_{\alpha },\Omega _{3}^{0}\right] 
\end{eqnarray*}%
and

\begin{eqnarray*}
\check{M}_{\alpha } &=&M_{\alpha \beta }n_{\beta },\text{ }\check{Q}^{\ast
}=Q_{\beta }^{\ast }n_{\beta },\text{ }\check{R}_{\alpha }=R_{\alpha \beta
}n_{\beta }, \\
\check{S}^{\ast } &=&S_{\beta }^{\ast }n_{\beta },\text{ }\check{N}_{\alpha
}=N_{\alpha \beta }n_{\beta },\check{M}^{\ast }=M_{\beta }^{\ast }n_{\beta },
\end{eqnarray*}%
In the above $n_{\beta }$ is the outward unit normal vector to $\Gamma _{u},$
and

\begin{eqnarray}
\Psi _{\alpha } &=&\frac{3}{h}\int_{-1}^{1}\zeta _{3}u_{\alpha }d\zeta _{3},
\notag \\
W &=&\frac{3}{4}\int_{-1}^{1}\left( 1-\zeta ^{2}\right) u_{3}d\zeta _{3}, 
\notag \\
\Omega _{\alpha }^{0} &=&\frac{3}{4}\int_{-1}^{1}\left( 1-\zeta ^{2}\right)
\varphi _{\alpha }d\zeta _{3}  \label{lagrange multipliers} \\
\Omega _{3} &=&\frac{3}{h}\int_{-1}^{1}\zeta _{3}\varphi _{3}d\zeta _{3} 
\notag \\
U_{\alpha } &=&\frac{1}{2}\int_{-1}^{1}u_{\alpha }d\zeta _{3},  \notag \\
\Omega _{3}^{0} &=&\frac{1}{2}\int_{-1}^{1}\varphi _{3}d\zeta _{3},  \notag
\end{eqnarray}%
Here $\Psi _{\alpha }$ are the rotations of the middle plane around $%
x_{\alpha }$ axis, $W$ \ the vertical deflection of the middle plate, $%
\Omega _{k}^{0}$ the microrotations in the middle plate around $x_{k}$ axis$%
, $ $U_{\alpha }$ is the in-plane displacements of the middle plane along $%
x_{a}$ axis$,$ $\Omega _{3}$ the rate of change of the microrotation $%
\varphi _{3}$ along $x_{3}$.

We also obtain the correspondence between the weighted displacement and the
microrotations (\ref{lagrange multipliers}) and the kinematic variables by
applying (\ref{kin 1}) and (\ref{kin 2}) in integration of expressions (\ref%
{lagrange multipliers}):

\begin{eqnarray}
\Psi _{\alpha } &=&V_{\alpha }(x_{1},x_{2},t),W=w(x_{1},x_{2},t),
\label{EQ correspondence} \\
\text{ }\Omega _{\alpha }^{0} &=&k_{1}^{\ast }\Theta _{\alpha
}^{0}(x_{1},x_{2},t),\text{ }\Omega _{3}=\frac{k_{2}^{\ast }}{h}\Theta
_{3}(x_{1},x_{2},t),  \notag \\
U_{\alpha } &=&U_{\alpha }(x_{1},x_{2},t),\text{ }\Omega _{3}^{0}=\Theta
_{3}^{0}(x_{1},x_{2},t),\   \notag
\end{eqnarray}%
where coefficients $k_{1\text{ }}^{\ast }$ and $k_{2\text{ }}^{\ast }$
depend on the variation of microrotations. Under the conditions (\ref{kin 2}%
) we have that $k_{1}^{\ast }=\frac{4}{5}$ and $k_{2}^{\ast }=\frac{8}{5}.$

The density of the work over the boundary $\Gamma _{\sigma }\times \left[
-h/2,h/2\right] $

\begin{equation*}
\mathcal{W}_{2}=\frac{h}{2}\int_{-1}^{1}\left( \sigma _{o\alpha }u_{\alpha
}+m_{o\alpha }\varphi _{\alpha }\right) n_{\alpha }d\zeta _{3}
\end{equation*}%
can be presented in the form \ \cite{Steinberg}%
\begin{equation*}
\mathcal{W}_{2}=\mathcal{S}_{o}\mathcal{\cdot U=}\Pi _{o\alpha }\Psi
_{\alpha }+\Pi _{o3}W+M_{o\alpha }\Omega _{\alpha }^{0}+M_{o3}^{\ast }\Omega
_{3}+\Sigma _{o,\alpha }U_{\alpha }+\Upsilon _{o3}\Omega _{3}^{0},
\end{equation*}%
where%
\begin{equation*}
\mathcal{S}_{o}\mathcal{=}\left[ \Pi _{o\alpha },\Pi _{o3},M_{o\alpha
},M_{o3}^{\ast },\Sigma _{o,\alpha },\Upsilon _{o3}\right]
\end{equation*}%
\begin{equation*}
\mathcal{U=}\left[ \Psi _{\alpha },W,\Omega _{\alpha }^{0},\Omega
_{3},U_{\alpha }\Omega _{3}^{0}\right] ,
\end{equation*}%
\begin{eqnarray}
M_{\alpha \beta }n_{\beta } &=&\Pi _{o\alpha },\ R_{\alpha \beta }n_{\beta
}=M_{o\alpha },  \notag \\
Q_{\alpha }^{\ast }n_{\alpha } &=&\Pi _{o3},\ S_{\alpha }^{\ast }n_{\alpha
}=M_{o3}^{\ast }.
\end{eqnarray}

\begin{eqnarray}
N_{\alpha \beta }n_{\beta } &=&\Sigma _{\alpha }, \\
M_{\alpha }^{\ast }n_{\alpha } &=&\Upsilon _{o3}.
\end{eqnarray}%
Now $n_{\beta }$ is the outward unit normal vector to $\Gamma _{\sigma },$
and 
\begin{eqnarray}
\Pi _{o\alpha } &=&\left( \frac{h}{2}\right) ^{2}\int_{-1}^{1}\zeta
_{3}\sigma _{o\alpha }d\zeta _{3},\ M_{o\alpha }=\frac{h}{2}\int_{-1}^{1}\mu
_{o\alpha }d\zeta _{3},  \notag \\
\Pi _{o3} &=&\frac{h}{2}\int_{-1}^{1}\left( \sigma _{o3}-\sigma _{0}\right)
d\zeta _{3},\ M_{o3}^{\ast }=\frac{h}{2}\int_{-1}^{1}(\mu
_{o3}-tn_{3})d\zeta _{3},  \notag \\
\Sigma _{o,\alpha } &=&\frac{h}{2}\int_{-1}^{1}\sigma _{o\alpha }d\zeta
_{3},\ \Upsilon _{o3}=\left( \frac{h}{2}\right) ^{2}\int_{-1}^{1}\zeta
_{3}\left( \mu _{o3}-\zeta _{3}v\right) d\zeta _{3}.
\end{eqnarray}%
We are able to evaluate the work done at the top and bottom of the Cosserat
plate by\ using boundary conditions (\ref{Bound conditions 0}) and (\ref%
{Bound conditions 1a})

\begin{equation*}
\int_{T\cup B}\left( \sigma _{o3}u_{3}+m_{o3}\varphi _{o3}\right)
n_{3}da=\int\limits_{P_{0}}(pW+v\Omega _{3}^{0})da.
\end{equation*}

\subsection{The Cosserat plate internal work density}

Here we define the density of the work done by the stress and couple stress
over the Cosserat strain field:

\begin{equation}
\mathcal{W}_{3}=\frac{h}{2}\int_{-1}^{1}\left( \mathbf{\sigma \cdot \gamma
+\mu \cdot \chi }\right) d\zeta _{3}.  \label{C1}
\end{equation}

Substituting stress and couple stress assumptions (\ref{stress assumption 1}%
) - (\ref{stress assump 3(a)}) and integrating expression (\ref{C1}) \ we
obtain the following expression \cite{Steinberg}:

\begin{equation}
\mathcal{W}_{3}=\mathcal{S\cdot E=}M_{\alpha \beta }e_{\alpha \beta
}+Q_{\alpha }\omega _{\alpha }+Q_{3\alpha }^{\ast }\omega _{\alpha }^{\ast
}+R_{\alpha \beta }\tau _{\alpha \beta }+S_{\alpha }^{\ast }\tau _{3\alpha
}+N_{\alpha \beta }\upsilon _{\alpha \beta }+M_{\alpha }^{\ast }\tau
_{3,\alpha }^{0},  \label{Work density 2}
\end{equation}%
where$\ \mathcal{E}$ is the Cosserat plate strain set of the the weighted
averages of strain and torsion tensors%
\begin{equation*}
\ \mathcal{E=}\left[ e_{\alpha \beta },\omega _{\beta },\omega _{a}^{\ast
},\tau _{3\alpha },\tau _{\alpha \beta },\upsilon _{\alpha \beta },\tau
_{3,\alpha }^{0}\right] .
\end{equation*}%
Here the components of\ $\mathcal{E}$ are 
\begin{eqnarray}
e_{\alpha \beta } &=&\frac{3}{h}\int_{-1}^{1}\zeta _{3}\gamma _{\alpha \beta
}d\zeta _{3},  \label{s1} \\
\omega _{\alpha } &=&\frac{3}{4}\int_{-1}^{1}\gamma _{\alpha 3}\left(
1-\zeta ^{2}\right) d\zeta _{3},  \label{s2} \\
\omega _{\alpha }^{\ast } &=&\frac{3}{4}\int_{-1}^{1}\gamma _{3\alpha
}\left( 1-\zeta ^{2}\right) d\zeta _{3},  \label{s3} \\
\tau _{3\alpha } &=&\frac{3}{h}\int_{-1}^{1}\zeta _{3}\chi _{3\alpha }d\zeta
_{3},  \label{s4} \\
\tau _{\alpha \beta }^{0} &=&\frac{3}{4}\int_{-1}^{1}\chi _{\alpha \beta
}\left( 1-\zeta ^{2}\right) d\zeta _{3},  \label{s5} \\
\upsilon _{\alpha \beta } &=&\frac{1}{2}\int_{-1}^{1}\gamma _{\alpha \beta
}d\zeta _{3},  \label{s6} \\
\tau _{3\alpha }^{0} &=&\frac{1}{2}\int_{-1}^{1}\chi _{3\alpha }d\zeta _{3}.
\label{s7}
\end{eqnarray}

The components of Cosserat plate strain (\ref{s1})-(\ref{s7}) can also be
represented in terms of the components of set $\mathcal{U}$ by the following
formulas \cite{Steinberg}:{\ }%
\begin{eqnarray}
e_{\alpha \beta } &=&\Psi _{\beta ,\alpha }+\varepsilon _{3\alpha \beta
}\Omega _{3},\text{ }  \notag \\
\omega _{\alpha } &=&\Psi _{\alpha }+\varepsilon _{3\alpha \beta }\Omega
_{\beta }^{0},  \notag \\
\omega _{\alpha }^{\ast } &=&W_{,\alpha }+\varepsilon _{3\alpha \beta
}\Omega _{\beta }^{0},  \notag \\
\tau _{3\alpha } &=&\Omega _{3,\alpha },  \label{strain -displ plate} \\
\tau _{\alpha \beta }^{0} &=&\Omega _{\beta ,\alpha }^{0},  \notag \\
\upsilon _{\alpha \beta } &=&U_{\beta ,\alpha }+\varepsilon _{3\alpha \beta
}\Omega _{3}^{0},  \notag \\
\tau _{3\alpha }^{0} &=&\Omega _{3,\alpha }^{0}.  \notag
\end{eqnarray}%
We call the relation\ (\ref{strain -displ plate}) the Cosserat plate
strain-displacement relation.

\subsection{The density of the kinetic energy}

Here we define the density of the kinetic energy:

\begin{equation*}
\mathcal{W}_{4}=\frac{h}{2}\int_{-1}^{1}\left( \mathbf{\rho \ddot{u}\cdot u+J%
}\text{ }\mathbf{\ddot{\varphi}\cdot \varphi }\right) d\zeta _{3},
\end{equation*}

which can be presented in the form

\begin{equation*}
\mathcal{W}_{4}=K\ddot{U}\cdot U\mathcal{=}I_{o}\ddot{\Psi}_{\alpha }\Psi
_{\alpha }+\rho _{o}\ddot{W}W+I_{o\alpha }\ddot{\Omega}_{\alpha }^{0}\Omega
_{\alpha }^{0}+J_{3}^{\ast }\ddot{\Omega}_{3}\Omega _{3}+\rho _{o}\ddot{U}%
_{\alpha }U_{\alpha }+I_{o3}\ddot{\Omega}_{3}^{0}\Omega _{3}^{0},
\end{equation*}%
where%
\begin{equation*}
U\mathcal{=}\left[ \hat{\Psi}_{\alpha },\ddot{W},\ddot{\Omega}_{\alpha }^{0},%
\ddot{\Omega}_{3},\ddot{U}_{\alpha },\ddot{\Omega}_{3}^{0}\right] ,
\end{equation*}

\begin{equation*}
K\mathcal{=}\left[ I_{o},\rho _{o},I_{o\alpha },J_{3}^{\ast },\rho
_{o},I_{o3}\right] ,
\end{equation*}%
and

\begin{equation*}
K\ddot{U}\mathcal{=}\left[ I_{o}\ddot{\Psi}_{\alpha },\rho _{o}\ddot{W}%
,I_{o\alpha }\ddot{\Omega}_{\alpha }^{0},J_{3}^{\ast }\ddot{\Omega}_{3},\rho
_{o}\ddot{U}_{\alpha },I_{o3}\ddot{\Omega}_{3}^{0}\right] ,
\end{equation*}%
where

\begin{eqnarray*}
I_{o} &=&\frac{\rho h^{3}}{12},\text{ }\rho _{o}=\rho h,\text{ }I_{o\alpha
}=k_{3}^{\ast }J_{a}h,\text{ }J_{3}^{\ast }=k_{4}^{\ast }J_{3}h^{3},\text{ }%
I_{o3}=J_{3}h, \\
k_{3}^{\ast } &=&\frac{5}{6},\text{ }k_{4}^{\ast }=\frac{25}{32}.
\end{eqnarray*}

\section{Cosserat Plate HPR Principle}

It is natural now to reformulate HPR variational principle for the Cosserat
plate $P$. Let $\mathcal{A}$ denote the set of all admissible states that
satisfy the Cosserat plate strain-displacement relation (\ref{strain -displ
plate}) and let $\Theta $ be a HPR\ functional on $\mathcal{A}$ defined \ by

\begin{equation}
\Theta ({\large s)}=U_{K}^{S}-\int\limits_{P_{0}}(\mathcal{S\cdot E-}K\ddot{U%
}\cdot U-pW-v\Omega _{3}^{0})da+\int_{\Gamma _{\sigma }}\mathcal{S}_{o}%
\mathcal{\cdot }\left( \mathcal{U-U}_{o}\right) ds+\int_{\Gamma _{u}}%
\mathcal{S}_{n}\mathcal{\cdot U}ds,  \label{free_energy}
\end{equation}%
for every ${\large s}=\left[ \mathcal{U},\mathcal{E},\mathcal{S}\right] \in 
\mathcal{A},$ then 
\begin{equation*}
\delta \Theta ({\large s})=0
\end{equation*}%
is equivalent to the following plate bending (A)\ and twisting (B)\ mixed
problems.

A. The flexural motions system of equations:

\begin{eqnarray}
M_{\alpha \beta ,\alpha }-Q_{\beta } &=&I_{o}\ddot{\Psi}_{\beta },
\label{equilibrium_equations 1_A} \\
Q_{a,\alpha }^{\ast }+p &=&\rho _{o}\ddot{W},
\label{equilibrium_equations 1_B} \\
R_{\alpha \beta ,\alpha }+\varepsilon _{3\beta \gamma }\left( Q_{\gamma
}^{\ast }-Q_{\gamma }\right) &=&I_{o\beta }\ddot{\Omega}_{\beta }^{0},
\label{equilibrium_equations 1_C} \\
S_{\alpha ,\alpha }^{\ast }+\epsilon _{3\beta \gamma }M_{\beta \gamma }
&=&J_{3}^{\ast }\ddot{\Omega}_{3},  \label{equilibrium_equations 1_D}
\end{eqnarray}%
with the resultant traction boundary conditions :

\begin{eqnarray}
M_{\alpha \beta }n_{\beta } &=&\Pi _{o\alpha },\ R_{\alpha \beta }n_{\beta
}=M_{o\alpha },  \label{bc_0} \\
Q_{\alpha }^{\ast }n_{\alpha } &=&\Pi _{o3},\ S_{\alpha }^{\ast }n_{\alpha
}=\Upsilon _{o3},  \label{bc_1}
\end{eqnarray}%
at the part $\Gamma _{\sigma }$ and he resultant displacement boundary
conditions

\begin{equation}
\Psi _{\alpha }=\Psi _{o\alpha },\text{ }W=W_{o},\text{ }\Omega _{\alpha
}^{0}=\Omega _{o\alpha }^{0},\text{ }\Omega _{3}=\Omega _{o3},  \label{bu_1}
\end{equation}%
at the part $\Gamma _{u}.$

The constitutive formulas: 
\begin{eqnarray}
e_{\alpha \alpha } &=&\frac{\partial {\large \Phi }}{\partial M_{\alpha
\alpha }}=\frac{12(\lambda +\mu )}{h^{3}\mu (3\lambda +2\mu )}M_{\alpha
\alpha }  \label{Const1-2} \\
&&-\left\vert \varepsilon _{\alpha \beta 3}\right\vert \frac{6\lambda }{%
h^{3}\mu (3\lambda +2\mu )}M_{\beta \beta }+\frac{3\lambda }{5h\mu (3\lambda
+2\mu )}\left( Q_{\beta ,\beta }^{\ast }\right) ,
\end{eqnarray}

\begin{equation*}
e_{\alpha \beta }=\frac{\partial {\large \Phi }}{\partial M_{\alpha \beta }}=%
\frac{3(\alpha +\mu )}{h^{3}\alpha \mu }M_{\alpha \beta }+\frac{3(\alpha
-\mu )}{h^{3}\alpha \mu }M_{\beta \alpha },\text{ }\alpha \neq \beta
\end{equation*}

\begin{eqnarray}
\omega _{\alpha } &=&\frac{\partial {\large \Phi }}{\partial Q_{\alpha }}=%
\frac{3(\alpha -\mu )}{10h\alpha \mu }Q_{\alpha }^{\ast }+\frac{3(\alpha
+\mu )}{10h\alpha \mu }Q_{\alpha }, \\
\omega _{\alpha }^{\ast } &=&\frac{\partial {\large \Phi }}{\partial
Q_{\alpha }^{\ast }}=\frac{3(\alpha -\mu )}{10h\alpha \mu }Q_{\alpha }+\frac{%
3(\alpha +\mu )}{10h\alpha \mu }Q_{\alpha }^{\ast },  \notag
\end{eqnarray}%
\begin{eqnarray}
\tau _{\alpha \alpha }^{0} &=&\frac{\partial {\large \Phi }}{\partial
R_{\alpha \alpha }}=\frac{6(\beta +\gamma )}{5h\gamma (3\beta +2\gamma )}%
R_{\alpha \alpha }-  \label{Const1-3} \\
&&\left\vert \varepsilon _{\alpha \beta 3}\right\vert \frac{3\beta }{%
5h\gamma (3\beta +2\gamma )}R_{\beta \beta }-\frac{\beta }{2\gamma (3\beta
+2\gamma )}t,  \notag \\
&&\tau _{\alpha \beta }^{0}=\frac{\partial {\large \Phi }}{\partial R_{\beta
\alpha }}=\frac{3(\epsilon -\gamma )}{10h\gamma \epsilon }R_{\alpha \beta }+%
\frac{3(\gamma +\epsilon )}{10h\gamma \epsilon }R_{\beta \alpha },\text{ }%
\alpha \neq \beta  \notag
\end{eqnarray}

\begin{equation}
\tau _{3\alpha }=\frac{\partial {\large \Phi }}{\partial S_{\alpha }^{\ast }}%
=\frac{3(\gamma +\epsilon )}{h^{3}\gamma \epsilon }S_{\alpha }^{\ast }.
\label{Const1-4}
\end{equation}

B.\ The extensional motions system of equations:

\begin{eqnarray}
N_{\alpha \beta ,\alpha } &=&\rho _{o}\ddot{U}_{\beta },
\label{equilibrium_equations 2 _A} \\
M_{\alpha ,\alpha }^{\ast }+\epsilon _{3\beta \gamma }N_{\beta \gamma }+v
&=&I_{o3}\ddot{\Omega}_{3}^{0},  \label{equilibrium_equations 2 _B}
\end{eqnarray}%
with the resultant traction boundary conditions at $\Gamma _{\sigma }$:

\begin{eqnarray}
N_{\alpha \beta }n_{\beta } &=&\Sigma _{\alpha },  \label{bc_2a} \\
M_{\alpha }^{\ast }n_{\alpha } &=&M_{o3}^{\ast },  \label{bc_2}
\end{eqnarray}%
and the resultant displacement boundary conditions at $\Gamma _{u}$:%
\begin{equation}
U_{\alpha }=U_{o\alpha },\text{ }\Omega _{3}^{0}=\Omega _{o3}^{0}.
\label{bu_2}
\end{equation}

The constitutive formulas:

\begin{eqnarray}
\omega _{\alpha \alpha } &=&\frac{\partial {\large \Phi }}{\partial
N_{\alpha \alpha }}=\frac{\lambda +\mu }{h\mu (3\lambda +2\mu )}N_{\alpha
\alpha }  \notag \\
&&-\frac{\lambda }{2h\mu (3\lambda +2\mu )}N_{(\alpha +1)(\alpha +1)}-\frac{%
\lambda }{2\mu (3\lambda +2\mu )}\sigma _{0},  \label{Const1-1} \\
\omega _{\alpha \beta } &=&\frac{\partial {\large \Phi }}{\partial N_{\alpha
\beta }}=\frac{\alpha +\mu }{4h\alpha \mu }N_{\alpha \beta }+\frac{\alpha
-\mu }{4h\alpha \mu }N_{\beta \alpha },\text{ }\alpha \neq \beta
\label{Const1-5} \\
\tau _{3\alpha }^{0} &=&\frac{\partial {\large \Phi }}{\partial M_{\alpha
}^{\ast }}=\frac{\gamma +\epsilon }{4h\gamma \epsilon }M_{\alpha }^{\ast }.
\label{Const1-6}
\end{eqnarray}%
We also represent the above constitutive relation in the compact form:%
\begin{equation*}
\mathcal{E=K}\left[ \mathcal{S}\right] =\mathcal{K\cdot S},
\end{equation*}%
where we call $\mathcal{K}$ the compliance Cosserat plate tensor.

Proof of the theorem.\ The variation of $\Theta ({\large s)}$%
\begin{eqnarray*}
\delta \Theta ({\large s)} &=&\int\limits_{P_{0}}\left\{ \left( \mathcal{K}%
\left[ \mathcal{S}\right] -\mathcal{E}\right) \cdot \delta \mathcal{S-}K%
\ddot{U}\cdot \delta U-\mathcal{S\delta E+}p\delta W+v\delta \Omega
_{3}^{0}\right\} da \\
&&+\int_{\Gamma _{\sigma }}\left\{ \delta \mathcal{S}_{o}\mathcal{\cdot }%
\left( \mathcal{U-U}_{o}\right) +\mathcal{S}_{o}\mathcal{\cdot \delta U}%
\right\} ds+\int_{\Gamma _{u}}\mathcal{S}_{n}\mathcal{\cdot \delta U}ds.
\end{eqnarray*}

We apply Green's theorem and integration by parts for $\mathcal{S}$ and $%
\mathcal{\delta U}$ $\ \mathcal{\cite{Gurtin}}$ to the expression:

\begin{eqnarray*}
\int\limits_{P_{0}}\mathcal{S}\cdot \delta \mathcal{E}da
&=&\int\limits_{\partial P_{0}}\mathcal{S}_{o}\delta \mathcal{\cdot U}\text{ 
}ds\mathcal{-}\int\limits_{P_{0}}\{\left( M_{\alpha \beta ,\alpha }-Q_{\beta
}-\right) \delta \Psi _{\beta }+Q_{\alpha ,\alpha }^{\ast }\delta W \\
&&+\left( R_{\alpha \beta ,\alpha }+\varepsilon _{3\beta \gamma }\left(
Q_{\gamma }^{\ast }-Q_{\gamma }\right) R_{\alpha \beta ,\alpha }\right)
\delta \Omega _{\beta }^{0} \\
&&+\left( S_{\alpha ,\alpha }^{\ast }+\epsilon _{3\beta \gamma }M_{\beta
\gamma }\right) \delta \Omega _{3}+N_{\alpha \beta ,\alpha }\delta U_{\beta }
\\
&&+\left( M_{\alpha ,\alpha }^{\ast }+\epsilon _{3\beta \gamma }N_{\beta
\gamma }\right) \delta \Omega _{3}^{0}\}da.
\end{eqnarray*}

Then based on the fact that $\mathcal{\delta U}$ and $\delta \mathcal{E}$
satisfy the Cosserat plate strain-displacement relation (\ref{strain -displ
plate}), we obtain

\begin{eqnarray*}
\delta \Theta ({\large s)} &=&\int\limits_{P_{0}}\left\{ \left( \mathcal{K}%
\left[ \mathcal{S}\right] -\mathcal{E}\right) \cdot \delta \mathcal{S}-%
\mathcal{S\delta E}\right\} da \\
&&+\int\limits_{P_{0}}\{\left( M_{\alpha \beta ,\alpha }-Q_{\beta }-I_{o}%
\ddot{\Psi}_{\beta }\right) \delta \Psi _{\beta }+\left( Q_{\alpha ,\alpha
}^{\ast }+p-\rho _{o}\ddot{W}\right) \delta W \\
&&+\left( R_{\alpha \beta ,\alpha }+\varepsilon _{3\beta \gamma }\left(
Q_{\gamma }^{\ast }-Q_{\gamma }\right) R_{\alpha \beta ,\alpha }-I_{o\beta }%
\ddot{\Omega}_{\beta }^{0}\right) \delta \Omega _{\beta }^{0} \\
&&+\left( S_{\alpha ,\alpha }^{\ast }+\epsilon _{3\beta \gamma }M_{\beta
\gamma }-J_{3}^{\ast }\ddot{\Omega}_{3}\right) \delta \Omega _{3}+\left(
N_{\alpha \beta ,\alpha }-\rho _{o}\ddot{U}_{\beta }\right) \delta U_{\beta }
\\
&&+\left( M_{\alpha ,\alpha }^{\ast }+\epsilon _{3\beta \gamma }N_{\beta
\gamma }+v-I_{o3}\ddot{\Omega}_{3}^{0}\right) \delta \Omega _{3}^{0}\}da \\
&&+\int_{\Gamma _{\sigma }}\delta \mathcal{S}_{o}\mathcal{\cdot }\left( 
\mathcal{U-U}_{o}\right) ds+\int_{\Gamma _{u}}(\mathcal{S}_{o}-\mathcal{S}%
_{n})\mathcal{\cdot \delta U}ds.
\end{eqnarray*}

If ${\large s}$ is a solution of the mixed problem, then%
\begin{equation*}
\delta \Theta ({\large s)=}0.
\end{equation*}

On the other hand, some extensions of the fundamental lemma of calculus of
variations \cite{Gurtin} together with the fact that $\mathcal{U}$ and $%
\mathcal{E}$ satisfy the Cosserat plate strain-displacement relation (\ref%
{strain -displ plate}) imply that $\mathcal{S}$ is a solution of the A and B
mixed problems.

\section{ Field Equations Governing Flexural and Extensional Motions in
terms of Kinematics Variables}

For the future consideration we represent the constitutive relations in the
following form$\footnote{%
In the following formulas a subindex $\alpha ^{\prime }=1$ if $\alpha =2$
and $\alpha ^{\prime }=2$ if $\alpha =1$}$

\begin{equation}
M_{\alpha \alpha }=D\left( \Psi _{\alpha ,\alpha }+\nu \Psi _{\alpha
^{\prime },\alpha ^{\prime }}\right) +\frac{\nu h^{2}}{10(1-\nu )}p,
\label{Cons 1a}
\end{equation}%
\begin{equation}
M_{\alpha ^{\prime }\alpha }=\frac{D}{2}\frac{(1+v)}{(1-N^{2})}\left( \Psi
_{\alpha ^{\prime },\alpha }+\Psi _{\alpha ,\alpha ^{\prime
}}+2N^{2}(-1)^{\alpha +1}\left( \Omega _{3}-\Psi _{\alpha ^{\prime },\alpha
}\right) \right)  \label{Cons 2a}
\end{equation}%
\begin{eqnarray}
R_{\alpha ^{\prime }\alpha } &=&\kappa _{2}^{2}Gh\left(
(l_{t}^{2}-2l_{b}^{2})\Omega _{\alpha ^{\prime },\alpha
}^{0}+2l_{b}^{2}\Omega _{\alpha ,\alpha ^{\prime }}^{0}\right) ,
\label{Cons 3} \\
R_{\alpha \alpha } &=&\kappa _{2}^{2}Ghl_{t}^{2}\left( \Omega _{\alpha
,\alpha }^{0}+(1-\Psi )\left( \Omega _{\alpha ,\alpha }^{0}+\Omega _{\alpha
^{\prime },\alpha ^{\prime }}^{0}\right) \right) +\frac{2Gl_{t}^{2}(1-\Psi )%
}{\Psi }t,  \notag
\end{eqnarray}

\begin{eqnarray}
Q_{\alpha } &=&\kappa _{1}^{2}\frac{Gh}{(1-N^{2})}\left( W_{,\alpha }+\Psi
_{\alpha }-2N^{2}\left( W_{,\alpha }+(-1)^{\alpha ^{\prime }}\Omega _{\alpha
^{\prime }}^{0}\right) \right) ,  \notag \\
Q_{\alpha }^{\ast } &=&\kappa _{1}^{2}\frac{Gh}{(1-N^{2})}\left( W_{,\alpha
}+\Psi _{\alpha }-2N^{2}\left( \Psi _{\alpha }+(-1)^{\alpha }\Omega _{\alpha
^{\prime }}^{0}\right) \right) ,  \label{Cons 4}
\end{eqnarray}%
\begin{equation}
S_{\alpha }^{\ast }=\frac{Gl_{t}^{2}(4l_{b}^{2}-l_{t}^{2})h^{3}}{12l_{b}^{2}}%
\Omega _{3,\alpha },  \label{Cons 5}
\end{equation}

\begin{equation}
N_{\alpha \alpha }=\frac{Eh}{(1-\nu ^{2})}\left( U_{a,\alpha }+\nu U_{\alpha
^{\prime },\alpha ^{\prime }}\right) +\frac{h\nu }{1-\nu }\sigma _{0},
\label{Cons 6}
\end{equation}%
\begin{equation}
N_{\alpha ^{\prime }\alpha }=\frac{Gh}{(1-N^{2})}\left( U_{\alpha ^{\prime
},\alpha }+U_{\alpha ,\alpha ^{\prime }}-2N^{2}\left( U_{\alpha ^{\prime
},\alpha }+(-1)^{\alpha }\Omega _{3}^{0}\right) \right) ,  \label{Cons 7}
\end{equation}

\begin{equation}
M_{\alpha }^{\ast }=\frac{Gl_{t}^{2}(4l_{b}^{2}-l_{t}^{2})h}{l_{b}^{2}}%
\Omega _{3,\alpha }^{0},  \label{Cons 8}
\end{equation}%
where we use the following technical constants \cite{Gauthier 1}, \cite%
{Lakes}: the Young's modulus $E=\frac{\mu (3\lambda +2\mu )}{\lambda +\mu },$
the Poisson's ratio$\ \nu =\frac{\lambda }{2(\lambda +\mu )},$ the shear
modulus $G=\frac{E}{2(1+\nu )},$ the flexural rigidity of the plate $D=\frac{%
Eh^{3}}{12(1-\nu ^{2})},$ the characteristic length for torsion $l_{t}=\sqrt{%
\frac{\gamma }{\mu }},$ the characteristic length for bending$\ l_{b}=\frac{1%
}{2}\sqrt{\frac{\gamma +\epsilon }{\mu }},$ the coupling\ number $N=\sqrt{%
\frac{\alpha }{\mu +\alpha }},$ the polar ratio $\Psi =\frac{2\gamma }{\beta
+2\gamma },$ $\kappa _{1}^{2}=\frac{5}{6}$ and $\kappa _{2}^{2}=\frac{5}{3}.$

Remark: The values of $\kappa _{1}$ and $\kappa _{2}$ depend on the form of
approximation. For instance, in the Mindlin's case of dynamics \cite{Mindlin}
the value of $\kappa _{1}^{2}=\frac{\pi }{12}$ , i.e. is slightly different.

After substitution (\ref{Cons 1a}) - (\ref{Cons 8}) \ into (\ref%
{equilibrium_equations 1_A}) - (\ref{equilibrium_equations 1_D}) and (\ref%
{equilibrium_equations 2 _A}) - (\ref{equilibrium_equations 2 _B}) we obtain
bending and twisting governing systems. We write system for the flexural
motions in the form:

\begin{equation}
\mathbf{L}\left( \mathbf{\partial }_{\mathbf{x}}\right) \mathbf{H-F=}\frac{%
\partial \mathbf{p}}{\partial t}\mathbf{,}\text{ }\mathbf{x\in }P_{0},
\label{Bending system}
\end{equation}%
where $\mathbf{L}\left( \mathbf{\partial }_{\mathbf{x}}\right) =\mathbf{L}%
\left( \frac{\partial }{\partial x_{a}}\right) ,$

\begin{equation*}
\mathbf{L}\left( \mathbf{\xi }\right) =\mathbf{L}\left( \mathbf{\xi }%
_{\alpha }\right) =\left[ 
\begin{array}{cccccc}
L_{11} & L_{12} & L_{13} & L_{14} & 0 & L_{16} \\ 
L_{12} & L_{22} & L_{23} & L_{24} & -L_{16} & 0 \\ 
-L_{13} & -L_{23} & L_{33} & 0 & L_{35} & L_{36} \\ 
-L_{14} & L_{24} & 0 & L_{44} & 0 & 0 \\ 
0 & L_{16} & -L_{35} & 0 & L_{55} & L_{56} \\ 
L_{16} & 0 & L_{36} & 0 & -L_{56} & L_{66}%
\end{array}%
\right] ,
\end{equation*}%
\begin{equation*}
{\mathbf{H}}^{T}=\left[ 
\begin{array}{cccccc}
\Psi _{1} & \Psi _{2} & W & \Omega _{3} & \Omega _{1}^{0} & \Omega _{2}^{0}%
\end{array}%
\right] ,
\end{equation*}

and

\begin{equation*}
{\mathbf{F}}^{T}=\left[ 
\begin{array}{cccccc}
F_{1} & F_{2} & F_{3} & F_{4} & F_{5} & F_{6}%
\end{array}%
\right] .
\end{equation*}%
In the above%
\begin{equation*}
\mathbf{p}^{T}\mathbf{=}\left[ 
\begin{array}{cccccc}
\frac{\rho h^{3}}{12}\frac{\partial \Psi _{1}}{\partial t} & \frac{\rho h^{3}%
}{12}\frac{\partial \Psi _{2}}{\partial t} & \rho h\frac{\partial W}{%
\partial t} & k_{4}J_{3}h^{3}\frac{\partial \Omega _{3}}{\partial t} & 
k_{3}J_{1}h\frac{\partial \Omega _{1}^{0}}{\partial t} & k_{3}J_{2}h\frac{%
\partial \Omega _{2}^{0}}{\partial t}%
\end{array}%
\right] ,
\end{equation*}

\begin{eqnarray*}
L_{11} &=&L_{11}(\xi _{1},\xi _{2})=k_{1}\xi _{1}^{2}+k_{2}\xi
_{2}^{2}-k_{3},\text{ }L_{22}=L_{11}(\xi _{2},\xi _{1}),L_{33}=k_{4}\Delta ,
\\
L_{44} &=&k_{5}\Delta -k_{6},\text{ }L_{55}=L_{55}(\xi _{1},\xi
_{2})=k_{7}\xi _{1}^{2}+k_{8}\xi _{2}^{2}-k_{9},\text{ }L_{66}=-L_{55}(\xi
_{2},\xi _{1}), \\
L_{12} &=&k_{10}\xi _{1}\xi _{2},\text{ }L_{13}=k_{11}\xi
_{1},L_{14}=k_{12}\xi _{2},L_{16}=k_{13},L_{23}=k_{11}\xi _{2}, \\
L_{24} &=&k_{12}\xi _{1},L_{35}=-k_{13}\xi _{2},\text{ }L_{36}=k_{13}\xi
_{1},L_{56}=k_{14}\xi _{1}\xi _{2},\Delta =\xi _{1}^{2}+\xi _{2}^{2}, \\
\text{ }F_{1} &=&-\frac{h^{2}\nu (1-N^{2})}{10(1-\nu )}\frac{\partial p}{%
\partial x_{1}},\text{ }F_{2}=-\frac{h^{2}\nu (1-N^{2})}{10(1-\nu )}\frac{%
\partial p}{\partial x_{2}}, \\
F_{3} &=&-(1-N^{2})p,F_{4}=0,\text{ }F_{5}=-\frac{5h(1-N^{2})}{6}(1-\Psi )%
\frac{\partial t}{\partial x_{1}}, \\
\text{ }F_{6} &=&\frac{5h(1-N^{2})}{6}(1-\Psi )\frac{\partial t}{\partial
x_{2}}
\end{eqnarray*}%
Here 
\begin{eqnarray*}
k_{1} &=&D(1-N^{2}),\text{ }k_{2}=\frac{D(1-\nu )}{2},\text{ }k_{3}=-\frac{%
5Gh}{6},k_{4}=\frac{5Gh}{6}, \\
k_{5} &=&\frac{D(1-\nu )l_{t}^{2}(4l_{b}^{2}-l_{t}^{2})(1-N^{2})}{2l_{b}^{2}}%
,k_{6}=2N^{2}D(1-\nu ), \\
k_{7} &=&\frac{5h(1-N^{2})Gl_{t}^{2}(2-\Psi )}{3},\text{ }k_{8}=\frac{%
10h(1-N^{2})Gl_{b}^{2}}{3},\text{ }k_{9}=\frac{10hGN^{2}}{3}, \\
k_{10} &=&\frac{D(1+\nu -2N^{2})}{2},k_{11}=\frac{5Gh(2N^{2}-1)}{6}%
,k_{12}=DN^{2}(1-\nu ), \\
k_{13} &=&\frac{5GhN^{2}}{3},\text{ }k_{14}=\frac{5h(1-N^{2})G\left(
l_{t}^{2}(2-\Psi )-2l_{b}^{2}\right) }{3}.
\end{eqnarray*}

The correspondent boundary and initial conditions are%
\begin{eqnarray}
\mathbf{T}{(\partial _{x}){\mathbf{H}}-{\mathbf{F}^{\ast }}}{=\mathbf{0}{%
\mathbf{,}}} &&\text{ }\mathbf{x\in }\Gamma _{\sigma },  \label{bending_bc}
\\
{{\mathbf{H}}-{\mathbf{H}}}_{o} &{=}&{{\mathbf{0,}}}\text{ }\mathbf{x\in }%
\Gamma _{u},
\end{eqnarray}

and 
\begin{eqnarray*}
{\mathbf{H(x,}}0{\mathbf{)}} &\mathbf{=H}&{}^{0} \\
\partial _{t}{\mathbf{H(x,}}0{\mathbf{)}} &\mathbf{=\hat{H}}&{}^{0}
\end{eqnarray*}

where differential operator $\mathbf{T}\left( \mathbf{\partial }_{\mathbf{x}%
}\right) =\mathbf{T}\left( \frac{\partial }{\partial x_{a}}\right) ,$

\begin{equation*}
\mathbf{T}\left( \mathbf{\xi }\right) =\mathbf{T}\left( \mathbf{\xi }%
_{\alpha }\right) =\left[ 
\begin{array}{cccccc}
T_{11} & T_{12} & 0 & T_{14} & 0 & 0 \\ 
T_{21} & T_{22} & 0 & T_{24} & 0 & 0 \\ 
T_{31} & T_{32} & T_{33} & 0 & 0 & T_{36} \\ 
0 & 0 & 0 & T_{44} & 0 & 0 \\ 
0 & 0 & 0 & 0 & T_{55} & T_{56} \\ 
0 & 0 & 0 & 0 & T_{65} & T_{66}%
\end{array}%
\right] ,
\end{equation*}%
and%
\begin{eqnarray*}
{\mathbf{(}F^{\ast })}^{T} &=&\left[ 
\begin{array}{cccccc}
F_{1}^{\ast } & F_{2}^{\ast } & F_{3}^{\ast } & F_{4}^{\ast } & F_{5}^{\ast }
& F_{6}^{\ast }%
\end{array}%
\right] , \\
{\mathbf{H}}_{o}^{T} &=&\left[ 
\begin{array}{cccccc}
\Psi _{o1} & \Psi _{o2} & W_{o} & \Omega _{o3} & \Omega _{o1}^{0} & \Omega
_{o2}^{0}%
\end{array}%
\right] .
\end{eqnarray*}%
In the above 
\begin{eqnarray*}
T_{11} &=&T_{1}(\xi _{1},\xi _{2}),T_{22}=T_{1}(\xi _{2},\xi _{1}),T_{1}(\xi
_{1},\xi _{2})=Dn_{1}\xi _{1}+\frac{D(1+\nu )}{2(1-N^{2})}n_{2}\xi _{2}, \\
T_{33} &=&\frac{5Gh}{6(1-N^{2})}\left( n_{1}\xi _{1}+n_{2}\xi _{2}\right)
,T_{44}=\frac{Gl_{t}^{2}(4l_{b}^{2}-l_{t}^{2})h^{3}}{12l_{b}^{2}}(n_{1}\xi
_{1}+n_{2}\xi _{2}), \\
T_{55} &=&\frac{5Gh}{3}(l_{t}^{2}n_{1}(2-\Psi )\xi _{1}+2l_{b}^{2}n_{2}\xi
_{2}),T_{66}=\frac{5Gh}{3}(2l_{b}^{2}n_{1}\xi _{1}+l_{t}^{2}(2-\Psi
)n_{2}\xi _{2}), \\
T_{12} &=&D\nu n_{1}\xi _{2}+\frac{D(1+\nu )(1-2N^{2})}{2(1-N^{2})}n_{2}\xi
_{1},T_{14}=\frac{D(1+\nu )N^{2}}{1-N^{2}}n_{2}, \\
T_{21} &=&D\nu n_{2}\xi _{1}+\frac{D(1+\nu )(1-2N^{2})}{2(1-N^{2})}n_{1}\xi
_{2},T_{24}=-\frac{D(1+\nu )N^{2}}{1-N^{2}}n_{1}, \\
T_{31} &=&\frac{5Gh(1-2N^{2})}{6(1-N^{2})}n_{1},T_{32}=\frac{5Gh(1-2N^{2})}{%
6(1-N^{2})}n_{2},T_{36}=\frac{5GhN^{2}}{3(1-N^{2})}(n_{1}-n_{2}), \\
T_{56} &=&T_{65}=T_{2}(\xi _{2},\xi _{1}),\text{ }T_{2}(\xi _{1},\xi _{2})=%
\frac{5Gh}{3}(l_{t}^{2}n_{1}(1-\Psi )\xi _{2}+(l_{t}^{2}-2l_{b}^{2})n_{2}\xi
_{1}), \\
F_{1}^{\ast } &=&-\frac{\nu h^{2}}{10(1-\nu )}n_{1}p-\Pi _{o1},F_{2}^{\ast
}=-\frac{\nu h^{2}}{10(1-\nu )}n_{2}p-\Pi _{o2}, \\
F_{3}^{\ast } &=&-\Pi _{o3},\text{ }F_{4}^{\ast }=-M_{o3}, \\
F_{5}^{\ast } &=&-\frac{2Gl_{t}^{2}(1-\Psi )}{\Psi }n_{1}t-M_{o1},F_{6}^{%
\ast }=-\frac{2Gl_{t}^{2}(1-\Psi )}{\Psi }n_{2}t-M_{o2}.
\end{eqnarray*}

The governing system for the extensional motions is%
\begin{equation}
\mathbf{\tilde{L}}\left( \mathbf{\partial }_{\mathbf{x}}\right) \mathbf{%
\tilde{H}-\tilde{F}}\mathbf{=}\frac{\partial \mathbf{\tilde{p}}}{\partial t}%
\mathbf{,}\text{ }\mathbf{x\in }P_{0},  \label{Twisting System}
\end{equation}%
where

\begin{equation*}
\mathbf{\tilde{L}}\left( \mathbf{\xi }\right) =\mathbf{\tilde{L}}\left( 
\mathbf{\xi }_{\alpha }\right) =\left[ 
\begin{array}{ccc}
\tilde{L}_{11} & \tilde{L}_{12} & \tilde{L}_{13} \\ 
\tilde{L}_{21} & \tilde{L}_{22} & \tilde{L}_{23} \\ 
\tilde{L}_{31} & \tilde{L}_{32} & \tilde{L}_{33}%
\end{array}%
\right]
\end{equation*}%
and 
\begin{eqnarray*}
{\mathbf{\tilde{F}}}^{T} &=&\left[ 
\begin{array}{ccc}
\tilde{F}_{1} & \tilde{F}_{2} & \tilde{F}_{3}%
\end{array}%
\right] , \\
{\mathbf{\tilde{H}}}^{T} &=&\left[ 
\begin{array}{ccc}
U_{1} & U_{2} & \Omega _{3}^{0}%
\end{array}%
\right] , \\
\mathbf{\tilde{p}}^{T} &=&\partial _{t}\left[ 
\begin{array}{ccc}
\rho _{0}\frac{\partial U_{1}}{\partial t} & \rho _{0}\frac{\partial U_{2}}{%
\partial t} & I_{o3}\frac{\partial \Omega _{3}^{0}}{\partial t}%
\end{array}%
\right]
\end{eqnarray*}

Here 
\begin{eqnarray*}
\tilde{T}_{11} &=&\kappa _{1}\xi _{1}^{2}+\kappa _{2}\xi _{2}^{2},\text{ }%
\tilde{T}_{12}=\kappa _{3}\xi _{1}\xi _{2},\tilde{T}_{13}=2\kappa _{4}\xi
_{2}, \\
\tilde{T}_{21} &=&\tilde{T}_{12},\tilde{T}_{22}=\tilde{T}_{11},\text{ }%
\tilde{T}_{23}=2\kappa _{4}\xi _{1}, \\
\tilde{T}_{31} &=&-\kappa _{4}\xi _{2},\tilde{T}_{32}=\kappa _{4}\xi _{1},%
\text{ }\tilde{T}_{33}=\kappa _{5}(\xi _{1}^{2}+\xi _{2}^{2})-\kappa _{2}, \\
\tilde{F}_{1}^{\ast } &=&-\frac{\nu \kappa _{1}}{2G}\frac{\partial \sigma
_{0}}{\partial x_{1}},\text{ }\tilde{F}_{2}^{\ast }=-\frac{\nu \kappa _{1}}{%
2G}\frac{\partial \sigma _{0}}{\partial x_{2}},\text{ }\tilde{F}_{3}^{\ast
}=-\frac{(1-N^{2})}{Gh}v, \\
\kappa _{1} &=&\frac{2(1-N^{2})}{1-\nu },\text{ }\kappa _{2}=2N^{2},\text{ }%
\kappa _{3}=1-\kappa _{1}=\frac{(1+\nu -2N^{2})}{(1-\nu )}, \\
\kappa _{4} &=&N^{2},\text{ }\kappa _{5}=\frac{%
l_{t}^{2}(4l_{b}^{2}-l_{t}^{2})(1-N^{2})}{2l_{b}^{2}}.
\end{eqnarray*}

The boundary and initial conditions for the extensional system has the
following form:

\begin{eqnarray}
\mathbf{\tilde{\mathbf{T}}}{(\partial _{x}){\mathbf{\tilde{H}}}-{\mathbf{%
\tilde{F}^{\ast }}}}{={\mathbf{0,}}} &&\text{ }\mathbf{x\in }\Gamma _{\sigma
}  \label{twisting_bc} \\
{{\mathbf{\tilde{H}}}-{\mathbf{\tilde{H}}}}_{o} &{=}&{{\mathbf{0,}}}\text{ }%
\mathbf{x\in }\Gamma _{u},
\end{eqnarray}%
and%
\begin{eqnarray*}
{\mathbf{\tilde{H}(x,}}0{\mathbf{)}}\mathbf{=\tilde{H}}^{0} && \\
\partial _{t}{\mathbf{\tilde{H}(x,}}0{\mathbf{)}}\mathbf{=\breve{H}}^{0} &&
\end{eqnarray*}

where differential operator $\tilde{\mathbf{T}}\left( \mathbf{\partial }_{%
\mathbf{x}}\right) =\tilde{\mathbf{T}}\left( \frac{\partial }{\partial x_{a}}%
\right) ,$

\begin{equation*}
\tilde{\mathbf{T}}\left( \mathbf{\xi }\right) = \tilde{\mathbf{T}}\left( 
\mathbf{\xi } _{\alpha }\right) =\left[ 
\begin{array}{ccc}
\tilde{T}_{11} & \tilde{T}_{12} & \tilde{T}_{13} \\ 
\tilde{T}_{21} & \tilde{T}_{22} & \tilde{T}_{23} \\ 
0 & 0 & \tilde{T}_{33}%
\end{array}%
\right] ,
\end{equation*}

\begin{eqnarray*}
\left( \mathbf{\tilde{F}}^{\ast }\right) ^{T} &=&\left[ 
\begin{array}{ccc}
\tilde{F_{1}}^{\ast } & \tilde{F_{2}}^{\ast } & \tilde{F_{3}}^{\ast }%
\end{array}%
\right] \\
\left( {{\mathbf{\tilde{H}}}}_{o}\right) ^{T} &=&\left[ 
\begin{array}{ccc}
U_{o1} & U_{o2} & \Omega _{o3}^{0}%
\end{array}%
\right] .
\end{eqnarray*}

In the above 
\begin{eqnarray*}
\tilde{T}_{11} &=&\frac{Ehn_{1}}{1-\nu ^{2}}\xi _{1}+\frac{Ghn_{2}}{1-N^{2}}%
\xi _{2},\ \tilde{T}_{12}=\frac{Eh\nu n_{1}}{1-\nu ^{2}}\xi _{2}+\frac{%
Ghn_{2}(1-2N^{2})}{1-N^{2}}\xi _{1}, \\
\tilde{T}_{13} &=&\frac{2N^{2}Ghn_{2}}{1-N^{2}},\ \tilde{T}_{21}=\frac{Eh\nu
n_{2}}{1-\nu ^{2}}\xi _{1}+\frac{Ghn_{1}(1-2N^{2})}{1-N^{2}}\xi _{2}, \\
\ \tilde{T}_{22} &=&\frac{Ehn_{2}}{1-\nu ^{2}}\xi _{1}+\frac{Ghn_{1}}{1-N^{2}%
}\xi _{1},\tilde{T}_{23}=-\frac{2N^{2}Ghn_{1}}{1-N^{2}}, \\
\ \tilde{T}_{33} &=&\frac{Gl_{t}^{2}(4l_{b}^{2}-l_{t}^{2})h}{l_{b}^{2}}(\xi
_{1}n_{1}+\xi _{2}n_{2}), \\
\tilde{F}_{1}^{\ast } &=&-\Sigma _{0,1}-\frac{h\nu n_{1}}{1-\nu }\sigma _{0},%
\tilde{F}_{2}^{\ast }=-\Sigma _{0,2}-\frac{h\nu n_{2}}{1-\nu }\sigma _{0},%
\tilde{F_{3}}^{\ast }=-M_{03}.
\end{eqnarray*}

\section{Conclusion}

We proposed a new mathematical model for dynamics of Cosserat elastic plates
based on Reissner-Mindlin's plate theory. The polynomial approximations of
the variation of couple stress and micropolar rotations in the thickness
direction allowed us to project Cosserat 3D\ Elasticity dynamics equations
into the dynamics equations in the middle plane of the plate. We generalized
fir the dynamics Hellinger-Prange -Reissner (HPR)\ principle to derive the
dynamics equations in the middle plane and constitutive relationships for
the plate. In terms of the kinematic variables, the total system of dynamic
equations describes the flexural (subsystem of 6 equations) and the
extensional (subsystem of 3 equations) motions of the plate.

\end{document}